\documentclass[aps,pra,showpacs,twocolumn]{revtex4}
\usepackage{amsmath, amssymb}
\usepackage{graphicx}
\usepackage{subfigure}

\renewcommand{\vec}[1]{\mathbf{#1}}

\DeclareMathOperator{\re}{Re}
\DeclareMathOperator{\im}{Im}
\DeclareMathOperator{\tr}{Tr}

\newtheorem{theorem}{Theorem}

\begin{document}

\title{Nonclassical Moments and their Measurement}

\author{E. Shchukin}
\email{evgeny.shchukin@uni-rostock.de}
\author{W. Vogel}
\email{werner.vogel@uni-rostock.de}
\affiliation{Arbeitsgruppe Quantenoptik, Fachbereich  Physik,
Universit\"at Rostock, D-18051 Rostock, Germany}

\begin{abstract}
  Practically applicable criteria for the nonclassicality of quantum
  states are formulated in terms of different types of moments. For
  this purpose the moments of the creation and annihilation operators,
  of two quadratures, and of a quadrature and the photon number
  operator turn out to be useful.  It is shown that all the required
  moments can be determined by homodyne correlation measurements.  An
  example of a nonclassical effect that is easily characterized by our
  methods is amplitude-squared squeezing.
\end{abstract}


\maketitle

\section{Introduction}

Nonclassical effects have attracted substantial interest during the
last decades. In particular due to the improvements of experimental
techniques in the field of quantum optics nonclassical quantum states
could be created in practice. After the first demonstration of photon
antibunching~\cite{prl-39-691} also sub-Poissonian photon
statistics~\cite{prl-51-384} and quadrature
squeezing~\cite{prl-55-2409} could be realized.

In view of new possibilities of measurement and characterization of
quantum states it became possible to more precisely consider the
characterization of nonclassical effects. The experimental
reconstruction of quantum states was demonstrated by balanced homodyne
tomography~\cite{prl-70-1244} and some other methods, for a review
see~\cite{welsch}. In principle this allows one to completely
characterize the quantum states of elementary quantum systems. On this
basis an old question receives new interest: What are the typical
signatures of nonclassical effects?

Besides balanced homodyne detection also homodyne correlation measurements
have been considered. In the particular case of a weak local oscillator it has
been shown that the method renders it possible to determine unusual types of
moments, such as normally-ordered moments composed of both a field quadrature
and the field intensity~\cite{prl-67-2450,pra-51-4160,prl-85-1855}. An
important advantage of these measurement techniques consists in the fact that
even for small quantum efficiencies the correlation properties of interest are
not smoothed out, as it is known in balanced homodyning~\cite{welsch}. For the following it
is of some interest that the experimental determination of homodyne
correlations in a multichannel device can be performed in
practice~\cite{prl-87-253601}.

The definition of nonclassicality that is widely accepted in quantum
optics is based on the existence of a well-behaved $P$-function. This
means that a state is considered to have a classical counterpart if
the $P$-function has the properties of a probability
measure~\cite{pr-140B-676,scr-T12-34}, for a nonclassical state it
fails to be interpreted as a probability. However, in addition to this
definition also states have been considered to be nonclassical, whose
mean photon number is small~\cite{scr-T12-34} or whose quantum
fluctuations are close to the vacuum-noise level~\cite{prl-84-1849}.
It is also interesting that nonclassical effects in weak measurements
have been considered to appear even for coherent states and for
thermal states of small photon
numbers~\cite{pla-329-184,pra-70-052115}. The latter example would
support both additional signatures of nonclassicality. However, the
coherent-state example is only consistent with the requirement of
the quantum noise being close to the vacuum level~\cite{prl-84-1849}, it does
not require a small photon number. We also note that nonclassical states having  negativities in the Wigner function are included in the class of nonclassical states we are dealing with.

Let us consider some recent developments of characterizing
nonclassical effects of a single-mode quantum state. The attempt was
made to formulate criteria that allow one to relate the
nonclassicality to observable quantities. One approach consists in the
use of characteristic functions of the quadrature
distributions~\cite{prl-84-1849}. The usefulness of the
nonclassicality criterion formulated in this way was successfully
demonstrated in an experiment~\cite{pra-65-033830}. Later on this
criterion turned out to be the lowest order of a hierarchy of
necessary and sufficient conditions for
nonclassicality~\cite{prl-89-283601}. This hierarchy is equivalent to
the failure of Bochner's old criterion for the existence of a
classical characteristic function~\cite{ma-108-378,k-fapt}, when it is
applied to the characteristic function of the $P$-function.

More recently the attempt was made to reformulate the notion of nonclassicality in a form which allows us to derive necessary and sufficient conditions in various representations~\cite{pra-71-011802}. For example, necessary and sufficient conditions in terms of quadrature moments have been obtained in this manner. These conditions in terms of moments are equivalent to those in terms of characteristic functions and to the failure of the $P$-function to be a probability distribution. The reformulation of the problem in terms of moments includes, as special cases, some
previously discussed conditions~\cite{pra-46-485,oc-95-109,pu-39-573}. 
The connection of nonclassicality criteria with the $17$th Hilbert problem has also been considered~\cite{prl-94-153601}. Last but not least, it has been
proposed to measure the nonclassicality of a single-mode quantum state
via the entanglement potential~\cite{prl-94-173602}. This is of
particular interest since it shows that the further investigation of
the nonclassicality of single-mode quantum states is of importance
even for applications that require entangled state.

The aim of the present paper consists in a more general formulation of
the conditions for the nonclassicality on the basis of observable
moments~\cite{pra-71-011802}. Three different types of criteria will
be analyzed in detail, which are formulated in terms of moments of
annihilation and creation operators, of two quadrature operators, and
of a quadrature and the number operator. In the latter case we will
also discuss the relation of the criteria to Artin's solution of the
$17$th Hilbert problem. We will show that all the considered moments
can be observed in a straightforward manner by homodyne correlation
measurements.  As a particular example for the usefulness of the methods
under study we consider the characterization and detection of
amplitude-squared squeezing~\cite{pra-36-3796}.

The paper is organized as follows. Necessary and sufficient
criteria for the nonclassicality are derived in terms of appropriately
chosen moments in Sec~II. In Sec.~III we propose some approaches for
the measurement of the moments needed in the new versions of
nonclassicality criteria. Special types of sufficient criteria for
nonclassicality are consider in Sec.~IV, with particular emphasis on
the characterization of amplitude-squared squeezing. In Sec.~V the
concept is illustrated for a special type of minimum-uncertainty
amplitude-squared squeezed states. A summary and some conclusions are
given in Sec.~VI.

\section{Nonclassicality Criteria}
\label{sec-II}

The density operator $\hat{\varrho}$ of any quantum state can be
written in the diagonal coherent-state representation~\cite{prl-10-277, pr-131-2766}
\begin{equation}\label{eq:P}
    \hat{\varrho} = \int P(\alpha) |\alpha\rangle\langle\alpha|\,
    d^2\alpha.
\end{equation}
A quantum state is said to be nonclassical if the corresponding
$P$-function \eqref{eq:P} fails to be interpreted as a probability
distribution on the complex plane \cite{pr-140B-676, scr-T12-34}.
Since the $P$-function may be highly singular, the criterion must be
reformulated in terms of measured quantities before it could be
applied for the interpretation of experiments.

In the following we will derive different versions of criteria for the
nonclassicality of a quantum state in terms of normally-ordered
moments of two appropriately chosen observables. It will be shown that
among the possibilities of choosing two operators that completely
describe the algebra of the harmonic oscillator there are at least two
choices that lead to necessary and sufficient conditions for
nonclassicality in terms of moments, for a brief consideration of one
of these two cases see~\cite{pra-71-011802}. Other choices of
operators, however, may only allow one to derive sufficient conditions
for the nonclassicality in terms of moments. The reason for this is
closely related to Artin's solution of the $17$th
Hilbert problem~\cite{positive-polynomials}.

We will start with a brief review of the reformulation of the
nonclassicality criteria in terms of characteristic
functions~\cite{prl-84-1849}. In this manner it became possible to
formulate a complete hierarchy of nonclassicality criteria that can be
related to experimental data~\cite{prl-89-283601}.  This approach will
also be needed as the basis for a rigorous formulation of criteria in terms of
moments.

\subsection{Characteristic Functions}

The characteristic function $\Phi(\beta)$ of $P(\alpha)$, that is its
two dimensional Fourier transform, is defined as
\begin{equation}
    \Phi(\beta) = \int P(\alpha) e^{\alpha \beta^* - \alpha^* \beta}\, d^2\alpha.
\end{equation}
It is easy to verify that it obeys the conditions
\begin{equation}\label{Pcf0}
    \Phi(0) = \tr(\hat{\varrho})=1, \quad \Phi(-\beta) = \Phi^{\ast}(\beta).
\end{equation}
Moreover, $\Phi(\beta)$ is a continuous function of $\beta$ for any
quantum state. Therefore it obeys all the requirements to apply the
following theorem introduced by Bochner~\cite{ma-108-378,k-fapt}:
\begin{theorem}[Bochner theorem]: The function $P(\alpha)$ is a
  probability distribution on the complex plane if and only if for any
  smooth function $f(\alpha)$ with compact support the following
  expression is nonnegative
\begin{equation}\label{eq:Phi}
    \iint \Phi(\alpha-\beta) f^*(\alpha) f(\beta) \, d^2\alpha \, d^2\beta \geqslant 0.
\end{equation}
\end{theorem}
The Bochner theorem can also be formulated in a discrete version by
replacing the integrations in Eq.~(\ref{eq:Phi}) with sums.

The necessary and sufficient conditions for $P(\alpha)$ being a
probability measure can thus be reformulated in terms of the
characteristic function. The relation~(\ref{eq:Phi}) is fulfilled, if and only if for any order $k=2, 3, \ldots,$ and for all complex $\beta_1, \ldots, \beta_k$ the determinants
\begin{equation}\label{det-def}
    D_{k} = D_k(\beta_1, \ldots, \beta_k) =
    \begin{vmatrix}
        1 & \Phi_{12} & \cdots &\Phi_{1k} \\
        \Phi^{\ast}_{12} & 1 & \cdots & \Phi_{2k} \\
        \hdotsfor{4} \\
        \Phi^{\ast}_{1k} & \Phi^{\ast}_{2k} & \cdots & 1 \\
    \end{vmatrix}
\end{equation}
obey the conditions
\begin{equation}\label{det}
    D_{k} \geqslant 0,
\end{equation}
where $\Phi_{ij} = \Phi(\beta_i-\beta_j)$. Consequently we arrive at the following theorem~\cite{prl-89-283601}:
\begin{theorem}: A quantum state is nonclassical, if and only if
there exist values $\beta_i$, ($i=1,\cdots,k$) for which at least one of the
determinants $D_k$ ($k=2,\cdots,\infty$) attains negative
values:
\begin{equation}\label{eq:D_k}
    D_{k} <0.
\end{equation}
\end{theorem}
We note that it is straightforward to relate the characteristic
functions $\Phi(\beta)$ to observable characteristic functions of
quadrature distributions, for more details
see~\cite{prl-84-1849,pra-65-033830,prl-89-283601}.

\subsection{Moments of $\hat{a}^\dagger$, $\hat{a}$}

In order to derive criteria for nonclassicality in terms of moments,
let us start with the normally-ordered expectation value of Hermitian
operators of the form~\cite{pra-71-011802,prl-94-153601}
\begin{equation}\label{eq:ff}
    \langle:\hat{f}^\dagger \hat{f}:\rangle = \int |f(\alpha)|^2 P(\alpha)\,d^2\alpha.
\end{equation}
We will consider only such functions $\hat{f} =
\hat{f}(\hat{a}^\dagger, \hat{a})$ of the creation and annihilation
operators, $\hat{a}^\dagger$ and $\hat{a}$, respectively, whose
normally-ordered form exists. From the equation above it immediately
follows that on a classical state the mean value \eqref{eq:ff} is
nonnegative for any operator $\hat{f}$. Hence the occurrence of
negative mean values
\begin{equation}\label{eq:ff2}
    \langle:\hat{f}^\dagger \hat{f}:\rangle < 0
\end{equation}
is a clear signature of the nonclassicality of the quantum state under consideration.

Let us first derive conditions for classicality in terms of the
moments $\langle\hat{a}^{\dagger k}\hat{a}^l\rangle$. The fact that
the mean value \eqref{eq:ff} is nonnegative for any polynomial
function
\begin{equation}
    \hat{f}(\hat{a}^\dagger, \hat{a}) = \sum^K_{k=0}\sum^L_{l=0} c_{kl} \hat{a}^{\dagger k} \hat{a}^l
\end{equation}
of $\hat{a}^\dagger$ and $\hat{a}$ leads to the nonnegativity of the
following quadratic form
\begin{equation}\label{eq:fff}
    \langle:\hat{f}^\dagger \hat{f}:\rangle = \sum^K_{n,k=0}\sum^L_{m,l=0} c^*_{kl} c_{nm}
    \langle \hat{a}^{\dagger n+l} \hat{a}^{m+k} \rangle,
\end{equation}
where the coefficients $c_{nm}$ are considered as independent variables. Due to Silvester's
criterion it is equivalent to
express this condition in terms of the  determinants
$d_N$ defined by
\begin{equation}\label{eq:daa}
    d_N =
   \underbrace{
    \begin{vmatrix}
        1 & \langle \hat{a} \rangle & \langle \hat{a}^\dagger \rangle & \langle \hat{a}^2 \rangle & \langle \hat{a}^\dagger\hat{a} \rangle & \langle \hat{a}^{\dagger 2} \rangle & \ldots \\
        \langle \hat{a}^\dagger \rangle & \langle \hat{a}^\dagger\hat{a}  \rangle &
        \langle \hat{a}^{\dagger 2} \rangle & \langle \hat{a}^\dagger \hat{a}^2 \rangle & \langle \hat{a}^{\dagger 2} \hat{a} \rangle & \langle \hat{a}^{\dagger 3} \rangle & \ldots \\
        \langle \hat{a} \rangle & \langle \hat{a}^2 \rangle &
        \langle \hat{a}^\dagger\hat{a} \rangle & \langle \hat{a}^3 \rangle & \langle \hat{a}^\dagger \hat{a}^2 \rangle & \langle \hat{a}^{\dagger 2} \hat{a} \rangle & \ldots \\
        \langle \hat{a}^{\dagger 2} \rangle & \langle \hat{a}^{\dagger 2} \hat{a} \rangle & \langle \hat{a}^{\dagger 3} \rangle & \langle \hat{a}^{\dagger 2}\hat{a}^2 \rangle & \langle \hat{a}^{\dagger 3} \hat{a} \rangle & \langle \hat{a}^{\dagger 4} \rangle & \ldots \\
        \langle \hat{a}^\dagger\hat{a} \rangle & \langle \hat{a}^\dagger \hat{a}^2 \rangle & \langle \hat{a}^{\dagger 2} \hat{a} \rangle & \langle \hat{a}^\dagger \hat{a}^3 \rangle & \langle \hat{a}^{\dagger 2}\hat{a}^2 \rangle & \langle \hat{a}^{\dagger 3} \hat{a} \rangle & \ldots \\
        \langle \hat{a}^2 \rangle & \langle \hat{a}^3 \rangle & \langle \hat{a}^\dagger \hat{a}^2 \rangle & \langle \hat{a}^4 \rangle & \langle \hat{a}^\dagger \hat{a}^3 \rangle & \langle \hat{a}^{\dagger 2}\hat{a}^2 \rangle & \ldots 
    \end{vmatrix}
    }_N .
\end{equation}
The conditions for classicality than read as
\begin{equation}\label{eq:daa-cl}
    d_N \geqslant 0.
\end{equation}
To this end, however, these conditions have only been demonstrated to be
necessary for the state to be classical. In order to show that they
are necessary and sufficient, we will make use of Bochner's theorem.

To apply the Bochner theorem to moments, let us introduce the
following operator $\hat{f}$ for any smooth function $f(\alpha)$ with
compact support:
\begin{equation}\label{eq:uf}
    \hat{f} = \int \underline{f}(\alpha) :\hat{D}(\alpha): \, d^2\alpha.
\end{equation}
Due to the properties of $f(\alpha)$ and the following expansion of the
normally-ordered displacement operator $:\hat{D}(\alpha):$,
\begin{equation}
    :\hat{D}(\alpha): = \sum_{k, l = 0}^{+\infty} \frac{\alpha^k (-\alpha^*)^l}{k!l!}
    \hat{a}^{\dagger k} \hat{a}^l,
\end{equation}
the operator \eqref{eq:uf} is correctly defined and its normally ordered form
\begin{equation}\label{eq:faa}
    \hat{f} = \sum^{+\infty}_{k, l = 0} c_{kl} \hat{a}^{\dagger k} \hat{a}^l
\end{equation}
exists.

The left hand side of the expression~\eqref{eq:Phi} appearing in the Bochner
theorem is nothing else but the mean value \eqref{eq:ff} for the operator
\eqref{eq:uf},
\begin{equation}
    \langle:\hat{f}^\dagger \hat{f}:\rangle = \iint \Phi(\alpha-\beta) \underline{f}^*(\alpha) \underline{f}(\beta)
    \, d^2\alpha \, d^2\beta.
\end{equation}
On the other hand the mean value $\langle:\hat{f}^\dagger \hat{f}:\rangle$ can be represented as
the following series
\begin{equation}\label{eq:ffi}
    \langle:\hat{f}^\dagger \hat{f}:\rangle = \sum^{+\infty}_{n,k,m,l=0} c^*_{kl} c_{nm}
    \langle\hat{a}^{\dagger n+l} \hat{a}^{m+k}\rangle.
\end{equation}
Suppose that all the determinants \eqref{eq:daa} are nonnegative.
Then all finite sums \eqref{eq:fff} are also nonnegative. As the
series \eqref{eq:ffi} converges, it can be approximated by finite sums
\eqref{eq:fff} and due to this it must be also nonnegative.
Eventually the Bochner theorem states that this is equivalent to the
nonnegativity of the $P$-function or the classicality of the state
under consideration.

Consequently we formulate another theorem for the nonclassicality of a
quantum state:
\begin{theorem}: A quantum state is nonclassical if and only if at
least one of the determinants $d_N$ violates the
condition~(\ref{eq:daa-cl}), that is
\begin{equation}
d_N < 0, \quad N= 3,4,\dots .
\end{equation}
\end{theorem}
Note that $d_2$ represent the incoherent part of the photon number,
\begin{equation}
d_2 =\langle \hat{a}^\dagger \hat{a} \rangle - \langle \hat{a}^\dagger \rangle
\langle \hat{a}
\rangle .
\end{equation}
Since this quantity is always nonnegative it yields no condition for
nonclassicality.

It is possible to formulate other (sufficient) conditions for
nonclassicality by considering subdeterminants of $d_N$. These
subdeterminants are obtained by any pairwise cancellation of such
lines and columns in $d_N$ that cross in a diagonal element of the
matrix. The negativity of any such subdeterminant is a sufficient
condition for nonclassicality. Criteria of this type may be useful for
characterizing the nonclassical properties of special quantum states.
Examples for such subdeterminants and for states that can be properly
characterized by the resulting sufficient conditions will be studied in
Sec.~IV.

\subsection{Moments of $\hat{x}_\varphi$, $\hat{p}_\varphi$}

Let us now consider two quadrature operators $\hat{x}_\varphi$ and
$\hat{p}_\varphi$.
As the quadrature operators are defined as linear
combinations of $\hat{a}$ and $\hat{a}^\dagger$, the latter can be
simply expressed in terms of the quadratures as
\begin{equation}
\hat{a}  = \frac{e^{i\varphi}}{2} ( \hat{x}_\varphi + i \hat{p}_\varphi), \quad \hat{a}^\dagger  = \frac{e^{-i\varphi}}{2} ( \hat{x}_\varphi - i
 \hat{p}_\varphi).
\end{equation}
One can reformulate the criteria for nonclassicality in terms of the
normally-ordered moments $\langle :\hat{x}^n_\varphi
\hat{p}^m_\varphi: \rangle$ of the quadratures, as has been considered
in~\cite{pra-71-011802}. We note that instead of using
$\hat{x}_\varphi$ and $\hat{p}_\varphi$ the criteria could also be
formulated with two arbitrary noncommuting quadratures
$\hat{x}_\varphi$ and $\hat{x}_{\varphi'}$, where $\varphi \not=
\varphi' \pm k \pi$ ($k= 0, 1, 2, \dots$). This more general form of
the criteria is simply obtained by replacing $\hat{p}_\varphi$ with
$\hat{x}_{\varphi'}$ in the criteria derived below.

In the following we will make use of the fact that any operator
$\hat{f}$ whose normally-ordered form exists can be written as a
normally-ordered power series with respect to $\hat{x}_\varphi$ and
$\hat{p}_\varphi$,
\begin{equation}\label{eq:fxp}
    \hat{f} = \sum^{+\infty}_{n, m = 0} \tilde{c}_{nm} :\hat{x}^n_\varphi \hat{p}^m_\varphi:.
\end{equation}
It is important that we use normally ordering here. An expansion of
$\hat{f}$ of the same structure, but the normally-ordered terms
$:\hat{x}^n_\varphi \hat{p}^m_\varphi:$ being replaced with
$\hat{x}^n_\varphi \hat{p}^m_\varphi$, may not exist.
From the representation \eqref{eq:fxp} it follows
that the state is classical if and only if all the
determinants $\tilde{d}_N$ defined by
\begin{equation}\label{eq:dxp}
    \tilde{d}_N =
    \underbrace{
    \begin{vmatrix}
        1 & \langle : \hat{x}_\varphi : \rangle & \langle : \hat{p}_\varphi : \rangle & \ldots \\
        \langle : \hat{x}_\varphi : \rangle & \langle : \hat{x}^2_\varphi : \rangle &
        \langle : \hat{x}_\varphi\hat{p}_\varphi : \rangle & \ldots \\
        \langle : \hat{p}_\varphi : \rangle & \langle : \hat{x}_\varphi\hat{p}_\varphi : \rangle &
        \langle : \hat{p}^2_\varphi : \rangle &\ldots \\
        \hdotsfor{4}
    \end{vmatrix}
    }_N
\end{equation}
are nonnegative:
\begin{equation}\label{eq:dxp-cl}
    \tilde{d}_N \geqslant 0.
\end{equation}

Consequently, the criteria for nonclassicality can be equivalently
formulated by the following theorem:
\begin{theorem}: A quantum state is nonclassical if and only if at
least one of the determinants $\tilde{d}_N$ violates the
condition~(\ref{eq:dxp-cl}), that is
\begin{equation}\label{cond-xp}
\tilde{d}_N < 0, \quad N= 2,3,\dots .
\end{equation}
\end{theorem}
In this version already the condition $\tilde{d}_2 < 0$ gives insight into the
nonclassicality of quantum states. It represents the condition for quadrature
squeezing,
\begin{equation}
\tilde{d}_2 = \langle : (\Delta \hat{x}_\varphi)^2 : \rangle < 0 .
\end{equation}
Note that from the nonclassicality conditions in Eq.~(\ref{cond-xp}) together
with~(\ref{eq:dxp}) we may derive further conditions by pairwise cancellation
of such lines and columns in the determinants that cross in the main diagonal.
For example one may formulate conditions such as
\begin{equation}\label{eq:s}
    s^{(2)}_\varphi =
    \begin{vmatrix}
        \langle{:\hat{x}^2_\varphi:}\rangle & \langle{:\hat{x}^2_\varphi\hat{p}_\varphi:} \rangle\\
        \langle{:\hat{x}^2_\varphi\hat{p}_\varphi:}\rangle & \langle {:\hat{x}^2_\varphi\hat{p}^2_\varphi:}\rangle
    \end{vmatrix}
    < 0.
\end{equation}
Conditions formulated in terms of such subdeterminants in some cases
may be more efficient to describe particular nonclassical effects as
the formal application of the hierarchy of the determinants $d_N$, for
an example see~\cite{pra-71-011802}.

\subsection{Moments of $\hat{x}_\varphi$, $\hat{n}$}

It is also possible to use other pairs of operators that together with the
unit operator generate the whole operator algebra of the harmonic oscillator.
Let us consider the position operator $\hat{x}_\varphi$ and the photon number operator
$\hat{n}$. The annihilation  and creation operators $\hat{a}$ and
$\hat{a}^\dagger$ are expressed in terms of the $\hat{x}_\varphi$ and $\hat{n}$ as follows
\begin{equation}\label{eq:axn}
    \hat{a} = \frac{1}{2}(\hat{x}_\varphi + [\hat{x}_\varphi, \hat{n}])e^{i \varphi}, \quad
    \hat{a}^\dagger = \frac{1}{2}(\hat{x}_\varphi - [\hat{x}_\varphi, \hat{n}])e^{-i \varphi}.
\end{equation}
In the present case, however, we need to use the commutation relation
$[\hat{a}, \hat{a}^\dagger] = 1$ to express the operators $\hat{a}$,
$\hat{a}^\dagger$ in terms of $\hat{x}_\varphi$, $\hat{n}$. Therefore it is no longer
possible to substitute the expressions \eqref{eq:axn} into the expansion
\eqref{eq:faa} and to rewrite the normally-ordered form \eqref{eq:faa} of the
operator $\hat{f}$ in terms of normally-ordered expressions in $\hat{x}_\varphi$ and
$\hat{n}$ as follows:
\begin{equation}
\label{eq:fxn}
    \hat{f} = \sum^{+\infty}_{k, l = 0} \tilde{\tilde{c}}_{kl} :\hat{x}^k_\varphi \hat{n}^l:.
\end{equation}
To make this more clear, already a linear term in~\eqref{eq:faa}, such as
$c_{01} \hat{a}$, has no representation in the form of Eq.~\eqref{eq:fxn}.

As a consequence of this fact one cannot conclude that the determinants that are similar to
\eqref{eq:daa} and \eqref{eq:dxp}, but with the moments $\langle
:\hat{x}^k_\varphi \hat{n}^l: \rangle$ instead of
$\langle(\hat{a}^\dagger)^k\hat{a}^l\rangle$ and $\langle :\hat{x}^k_\varphi
\hat{p}^l_\varphi: \rangle$, respectively, form a complete hierarchy of
criteria. The reason for this is the following. The not-existence of a
representation of the form~\eqref{eq:fxn} for a general operator prevents one
from relating these determinants to the Bochner theorem. Thus the
corresponding determinants do not lead to necessary and sufficient conditions
for nonclassicality.

Because of this situation we will formulate only necessary conditions for
classicality or, the other way around, we will derive sufficient conditions
for nonclassicality in terms of the moments $\langle :\hat{x}^k_\varphi
\hat{n}^l: \rangle$.  The mean value of the normally-ordered operator
$:\hat{F}(\hat{x}_\varphi, \hat{n}):$ can be expressed in terms of the
$P$-function as
\begin{equation}\label{eq:F}
    \langle :\hat{F}(\hat{x}_\varphi, \hat{n}): \rangle = \int F(2\re(\alpha e^{-i \varphi}), |\alpha|^2)P(\alpha)\,d^2\alpha.
\end{equation}
For any classical state and for any operator $F(\hat{x}_\varphi,
\hat{n})$ with a nonnegative associated c-number function $F(2\re\alpha,
|\alpha|^2)$ the mean value \eqref{eq:F} is nonnegative.  If we
identify $:\hat{F}(\hat{x}, \hat{n}):$ with $:\hat{f}^\dagger \hat{f}:$
together with the special form $\hat{f}$ given in Eq.~\eqref{eq:fxn},
we derive necessary conditions for classicality in terms of the
determinants
\begin{equation}\label{eq:d1}
    d^{(1)}_N =
    \underbrace{
    \begin{vmatrix}
        1 & \langle : \hat{x}_\varphi : \rangle & \langle : \hat{n} : \rangle & \ldots \\
        \langle : \hat{x}_\varphi : \rangle & \langle : \hat{x}^2_\varphi : \rangle &
        \langle : \hat{x}_\varphi\hat{n} : \rangle & \ldots \\
        \langle : \hat{n} : \rangle & \langle : \hat{x}_\varphi\hat{n} : \rangle &
        \langle : \hat{n}^2 : \rangle &\ldots \\
        \hdotsfor{4}
    \end{vmatrix}
    }_N
\end{equation}
as
\begin{equation}\label{eq:d1a}
    d^{(1)}_N \geqslant 0.
\end{equation}
Consequently, if
\begin{equation}
 d^{(1)}_N  <0 , \quad N=2,3, \dots,
\end{equation}
the nonclassical properties of the considered quantum state have been verified
by a sufficient but not necessary condition.

The latter can be illustrated as follows. There exist operators $F(\hat{x}_\varphi,
\hat{n})$ of different kind whose associated $c$-number function
$\hat{F}(2\re(\alpha e^{i \varphi}), |\alpha|^2)$ is nonnegative. Let us consider
\begin{equation}
    F(\hat{x}_\varphi, \hat{n}) = (4\hat{n} - \hat{x}^2_\varphi) \hat{f}^\dagger(\hat{x}_\varphi, \hat{n})
    \hat{f}(\hat{x}_\varphi, \hat{n}).
\end{equation}
It is clear that the corresponding function is nonnegative
\begin{equation}
\begin{split}
    F(2\re(\alpha e^{-i\varphi}), |\alpha|^2) &= 4 \im^2 \left( \alpha
    e^{-i \varphi} \right)  \\
    &\times |f(2\re(\alpha e^{-i \varphi}),
    |\alpha|^2)|^2 \geqslant 0.
\end{split}
\end{equation}
This leads to the following necessary conditions for
the nonnegativity of the $P$-function:
\begin{equation}\label{eq:d2}
    d^{(2)}_N
    \geqslant 0,
\end{equation}
which is formulated in terms of the determinants
\begin{equation}
    d^{(2)}_N =
    \underbrace{
    \begin{vmatrix}
        4m_{01}-m_{20} & 4m_{02}-m_{21} & \ldots \\
        4m_{02}-m_{21} & 4m_{03}-m_{22} & \ldots \\
        \hdotsfor{3}
    \end{vmatrix}
    }_N
\end{equation}
with
\begin{equation}
    m_{kl} = \langle : \hat{x}^k_\varphi\hat{n}^l : \rangle.
\end{equation}

Consequently, the violation of the nonnegativity of one such determinant,
\begin{equation}
 d^{(2)}_N  <0 , \quad N= 1,2, \dots,
\end{equation}
is sufficient to demonstrate the nonclassical behavior of a quantum state.
Note that these conditions already characterize nonclassical effects
by the first order determinant $d_1$. By inspection of its expression,
\begin{equation}
    d_1 = 4\langle\hat{n}\rangle - \langle:\hat{x}^2_\varphi:\rangle = \langle:\hat{p}^2_\varphi:\rangle = \langle:\hat{x}^2_{\varphi+\pi/2}:\rangle,
\end{equation}
we observe that the condition $ d^{(2)}_1  <0$ reproduces the
condition for quadrature squeezing provided that
$\langle\hat{x}_{\varphi+\pi/2}\rangle =0$, that is for quantum
states whose mean amplitude is vanishing, as for example in a squeezed
vacuum state.

\subsection{Relation to the 17th Hilbert problem}

In his famous address to the 1900 International Congress of Mathematicians
Hilbert formulated 23 problems that he considered to be the
most important problems to be solved in the following century.
The 17th problem, in its simplest form, was solved by Artin in
1926. Artin's solution can be formulated in the following
theorem~(see e.g.~\cite{positive-polynomials}):
\begin{theorem}[$17$th Hilbert problem]: Any nonne\-ga\-tive polynomial $F(x_1, \ldots, x_n) \in \mathbf{R}[x_1, \ldots, x_n]$,
\begin{equation}
    F(x_1, \ldots, x_n) \geqslant 0, \quad \forall (x_1, \ldots, x_n) \in \mathbf{R}^n,
\end{equation}
can be represented as a finite sum of squares of rational functions
\begin{equation}
    F(x_1, \ldots, x_n) = \sum^N_{k = 1} R^2_k(x_1, \ldots, x_n),
\end{equation}
where $R_k(x_1, \ldots, x_n) \in \mathbf{R}(x_1, \ldots, x_n)$, i. e.
\begin{equation}
    R_k(x_1, \ldots, x_n) = \frac{F_k(x_1, \ldots, x_n)}{G_k(x_1, \ldots, x_n)},
\end{equation}
and all $F_k$ and $G_k$, $k = 1, \ldots, N$ are polynomials
\begin{equation}
    F_k(x_1, \ldots, x_n),\ G_k(x_1, \ldots, x_n) \in \mathbf{R}[x_1, \ldots, x_n].
\end{equation}
\end{theorem}

This theorem is useful for getting deeper insight in the general
problem of formulating conditions for the nonclassicality of quantum
states in such cases as discussed in the preceding subsection.
For this purpose let us consider the nonnegativity condition
to be fulfilled for any quantum state that has a classical
counterpart. Let us restrict the class of functions $F(x_1,x_2)$,
with $x_1=2\re(\alpha e^{-i \varphi})$ and $x_2=|\alpha|^2$, on the r.h.s. of
Eq.~\eqref{eq:F} to polynomials. Now we may formulate the nonclassicality
conditions in terms of the operators $\hat{x}_\varphi$ and $\hat{n}$
by using Artin's solution of the 17th Hilbert problem as follows:
\begin{equation}\label{noncl-xn}
 \langle : \hat{F} (\hat{x}_\varphi,\hat{n}):\rangle = \sum^N_{k = 1} \Big \langle
 :
 \frac{\hat{F}^2_k(\hat{x}_\varphi,\hat{n})}{\hat{G}^2_k(\hat{x}_\varphi,\hat{n})}:
\Big \rangle <0.
\end{equation}
That is, the condition for nonclassicality leads to
normally-ordered expectation values of fractions of squares of
polynomials. The extension of the conditions beyond polynomial functions is
not obvious, nor it is straightforward to formulate necessary and sufficient
conditions for nonclassicality in terms of normally-ordered moments of
$\hat{x}_\varphi$ and $\hat{n}$. Also other types of operators, even if they
render it possible to span the algebra of the harmonic oscillator by using
commutation rules, are expected to lead to similar problems.

In the cases of the operators $\hat{a}^\dagger$ and $\hat{a}$ on one hand and
$\hat{x}_\varphi$ and $\hat{p}_\varphi$ on the other hand these problems do
not exist. The general form of functions according to the solution of the 17.
Hilbert problem is not needed due to the direct relation of the considered
expectation values $\langle :\hat{f}^\dagger \hat{f}:\rangle$ to the Bochner
theorem.  This allowed us to directly formulate necessary and sufficient
conditions for the nonclassicality in terms of the moments of these operator.
In the following we will consider the possibilities to measure these moments.
This eventually allows one to characterize nonclassical states by measurable
moments.

\section{Measurement of moments}
\label{sec-III}

In the preceding section we have considered the formulation of
criteria for the nonclassicality of quantum states in terms of
moments. Such criteria are only of practical interest if one can
design measurement principles that allow one to determine the
corresponding moments. Since these schemes will depend on the types of
moments, we will consider them separately. All these detection schemes
will be based on homodyne correlation measurements. An important
advantage of such measurements consists in the fact that the
accessible information on a quantum state is not smoothed out be
small detection efficiencies.

For the first time homodyne correlation measurements with a weak local
oscillator have been proposed for measuring squeezing and anomalous
moments, containing nonequal powers in the annihilation and creation
operators, in resonance fluorescence~\cite{prl-67-2450}.  This
measurement principle was studied in more detail~\cite{pra-51-4160},
where the detection of quantum correlations of the photon number and a
quadrature operator has been analyzed. Later on such measurements have
also been considered for determining particular nonclassical
properties of light~\cite{prl-85-1855}. In the following we will
modify the method in such a way that it becomes possible to determine
the different types of moments considered in the preceding section.

\subsection{Detection of $\langle \hat{a}^{\dagger k} \hat{a}^l \rangle$}
\label{subsec-IIIa}

\begin{figure}
    \includegraphics{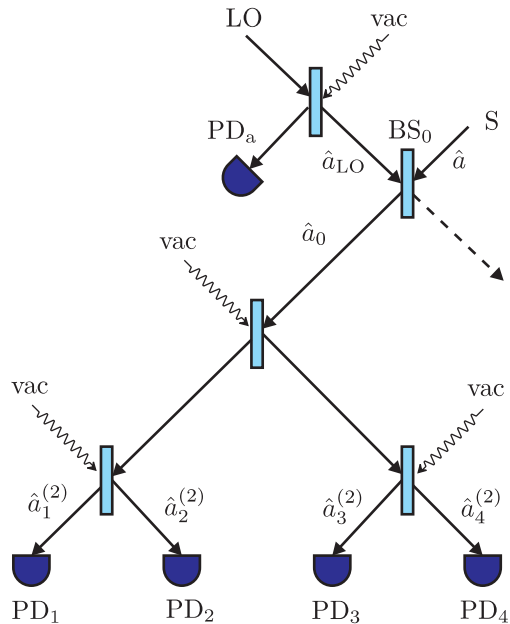}
\caption{Scheme for measuring the moments
$\langle(\hat{a}^\dagger)^k\hat{a}^l\rangle$.}\label{fig:moments}
\end{figure}

Let us consider the detection scheme presented in
Fig.~\ref{fig:moments}.  As an example we have shown a measurement
scheme of depth $d=2$, where the depth is the number of beam splitters
between the entrance beam splitter $\mathrm{BS}_0$ and any of the
detectors $\mathrm{PD}_1,\dots \mathrm{PD}_4$. An auxiliary
photodetector $\mathrm{PD}_{\mathrm{a}}$ is used here and in the
following measurement schemes to record, simultaneously with the
correlation measurements of interest, the intensity of the local
oscillator.  In principle the measurement device could be further
extended to an arbitrary value of depth $d$. Below we will show that
such a scheme allows one to measure the moments $\langle
\hat{a}^{\dagger k} \hat{a}^l \rangle$ for $k, l = 0, 1, \ldots, 2^d$.

The entrance beam-splitter for the signal field, $\mathrm{BS_0}$, is characterized by the
parameters $T_0$ and $R_0$ with
\begin{equation}
    |T_0| \gg |R_0|,
\end{equation}
so that the signal field of interest is almost completely used for the
correlation measurements. In our scheme the strength of the local
oscillator on the detectors is typically of the same order of
magnitude as the signal field, which is easily realized with a weakly
reflecting entrance beam splitter.  All the other beam splitters in the
device are $50$\%-$50$\%, so that their parameters $T_k$ and $R_k$ can be
written in the simple form
\begin{equation}
    T_k = \frac{1}{\sqrt{2}}e^{i\varphi_k}, \quad R_k = \frac{i}{\sqrt{2}}e^{i\varphi_k}.
\end{equation}
The operator $\hat{a}_0$ describing the field behind the entrance beam
splitter $\mathrm{BS_0}$ reads as
\begin{equation}
    \hat{a}_0 = \frac{1}{\sqrt{2}}(T_0 \hat{a} + R_0 \hat{a}_{\mathrm{LO}}).
\end{equation}
It is clear that all the operators $\hat{a}^{(d)}_k$, that describe
the fields on the $k$-th detector, are proportional to the operator $\hat{a}_0$,
\begin{equation} \label{eq:det-a}
    \hat{a}^{(d)}_k = \frac{e^{i\Phi_k}}{\sqrt{2^d}}\hat{a}_0 + \ldots,
\end{equation}
where $\ldots$ stands for a combination of the corresponding vacuum-channel operators which gives no contribution to the quantities considered below. The phase $\Phi_k$ depends on the path leading from the beam-splitter $\mathrm{BS_0}$ to the
$k$-th photodetector $\mathrm{PD_k}$.

The measurement scheme can be used to detect the coincidences
registered by all $n$ photodetectors, which are described by the
normally-ordered correlation functions $\Gamma_{k_1, \ldots, k_n}$ of the form
\begin{equation}
    \Gamma_{k_1, \ldots, k_n} = \langle : \hat{a}^{(d)\dagger}_{k_1}\hat{a}_{k_1}^{(d)} \ldots
    \hat{a}^{(d)\dagger}_{k_n}\hat{a}_{k_n}^{(d)} : \rangle.
\end{equation}
Using Eq.~\eqref{eq:det-a} it is clear that these correlations do not
depend on $k_1, \ldots, k_n$ and they can be written as
\begin{equation}
    \Gamma^{(n)} = \Gamma_{k_1, \ldots, k_n} =
    \frac{1}{2^{n d}}\langle \hat{a}_0^{\dagger n} \hat{a}_0^n \rangle.
\end{equation}
Summing up over all possible combinations of $n$ photo-detectors, in
order to avoid loss of measured data, we get the following function:
\begin{equation}\label{eq:Fn}
\begin{split}
    F_n &= \sum_{\{k_1, \ldots, k_n\}}\Gamma_{k_1, \ldots, k_n} = \\
    &\binom{2^d}{n} \Gamma^{(n)} = \sum^n_{m=-n} f_n(m) e^{i m \varphi},
\end{split}
\end{equation}
where $\varphi = \varphi_{\mathrm{LO}} - \pi/2$ and the coefficients $f_n(m)$ read
\begin{equation}\label{eq:f}
\begin{split}
    f_n(m) = \frac{\binom{2^d}{n}}{2^{nd}} &\sum_{k-l = m} \binom{n}{k} \binom{n}{l} \times \\
    &|T_0|^{k+l} |R_0 \alpha|^{2n-k-l} \langle \hat{a}^{\dagger k} \hat{a}^l \rangle.
\end{split}
\end{equation}
The function $F_n$ in Eq.~\eqref{eq:Fn} is a function of the phase $\varphi$
\begin{equation}
    F_n = F_n(\varphi),
\end{equation}
and the coefficients $f_n(m)$ can be obtained from this function using Fourier
transform
\begin{equation}
    f_n(m) = \frac{1}{2\pi}\int_0^{2\pi} F_n(\varphi) e^{-i m \varphi}\,d\varphi.
\end{equation}

Each coefficient $f_n(m)$ in Eq.~\eqref{eq:f} is a combination of some moments $\langle \hat{a}^{\dagger
k} \hat{a}^l \rangle$. From these combinations it is possible to extract the moments $\langle
\hat{a}^{\dagger k} \hat{a}^l \rangle$ themselves step by step. The moments
$\langle\hat{a}\rangle$, $\langle\hat{a}^\dagger\rangle$ and $\langle\hat{a}^\dagger\hat{a}\rangle$
can be obtained directly from $F_1(\varphi)$,
\begin{equation}
\begin{split}
    \langle\hat{a}\rangle &= \frac{f_1(-1)}{|T_0 R_0 \alpha|}, \\
    \langle\hat{a}^\dagger\rangle &= \frac{f_1(1)}{|T_0 R_0 \alpha|}, \\
    \langle\hat{a}^\dagger\hat{a}\rangle &= \frac{f_1(0)-|R_0 \alpha|^2}{|T_0|^2}.
\end{split}
\end{equation}
Using these moments the next step gives the explicit expressions for the moments $\langle
\hat{a}^{\dagger k} \hat{a}^l \rangle$, $k, l = 1, 2$. We present the expressions only for the
moments $\langle\hat{a}^2\rangle$ and $\langle\hat{a}^\dagger\hat{a}^2\rangle$,
\begin{equation}
\begin{split}
    \langle\hat{a}^2\rangle &= \frac{8}{3} \frac{f_2(-2)}{|T_0 R_0 \alpha|^2}, \\
    \langle\hat{a}^\dagger\hat{a}^2\rangle &= \frac{1}{3}
    \frac{4f_2(-1)-3|R_0 \alpha|^2f_1(-1)}{|T^3_0 R_0 \alpha|}.
\end{split}
\end{equation}

\begin{figure}
    \includegraphics{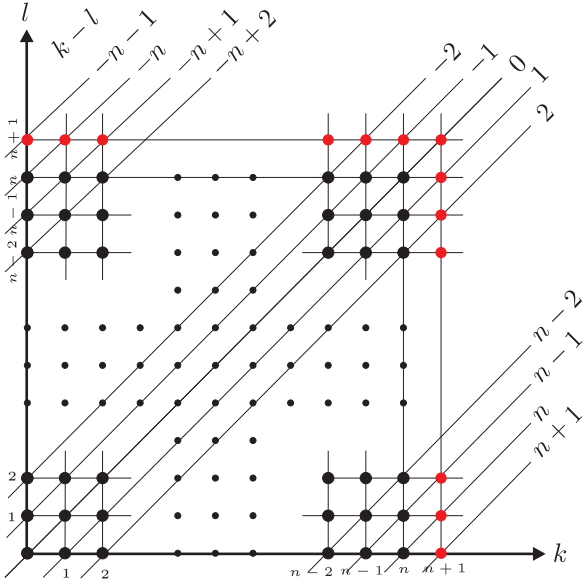}
\caption{Step by step extraction of the moments $\langle \hat{a}^{\dagger k} \hat{a}^l \rangle$.}
\label{fig:kl}
\end{figure}

Figure~\ref{fig:kl} illustrates the possibility to extract the general moments $\langle
\hat{a}^{\dagger k} \hat{a}^l \rangle$, $0 \leqslant k, l \leqslant 2^d$ step by step. The $n$-th
step is the extraction of the moments $\langle \hat{a}^{\dagger k} \hat{a}^l \rangle$ with $0
\leqslant k, l \leqslant n$, i. e. the moments $\langle \hat{a}^{\dagger k} \hat{a}^l \rangle$ with
$(k, l)$ corresponding to the square
\begin{equation}
    S_n = \{(k, l) \in \mathbf{Z}^2| 0 \leqslant k, l \leqslant n\}.
\end{equation}
But, in fact, we don't measure the moments $\langle \hat{a}^{\dagger k} \hat{a}^l \rangle$
themselves, we measure their combinations $f_n(m)$, $m = -n, \ldots, n$, that consist of the
moments $\langle \hat{a}^{\dagger k} \hat{a}^l \rangle$, $0 \leqslant k, l \leqslant n$ with
additional condition $k-l=m$. Geometrically the combination $f_n(m)$ contains the moments $\langle
\hat{a}^{\dagger k} \hat{a}^l \rangle$ with $(k, l)$ in the intersection of the square $S_n$ and
the inclined line $L_m: k-l=m$. This means that using the data obtained on one step only it is
impossible to extract the moments $\langle \hat{a}^{\dagger k} \hat{a}^l \rangle$, but assuming we
have the data of previous steps it is: each combination $f_{n+1}(m)$ contains one new moment by
comparison with $f_n(m)$. Note that $f_{n+1}(n+1)$ and $f_{n+1}(-n-1)$ contain one moment only,
$\langle \hat{a}^{\dagger n+1} \rangle$ and $\langle \hat{a}^{n+1} \rangle$
correspondingly, so that these moments can be measured without keeping the information of previous
steps. Geometrically this means that $S_{n+1} \cap L_m$ contains one point more by comparison with
$S_n \cap L_m$.  So, step by step it is possible to extract all the moments $\langle
\hat{a}^{\dagger k} \hat{a}^l \rangle$, $0 \leqslant k, l \leqslant 2^d$. The last step, the
$2^d$-th, consists in the extraction of the moments $\langle \hat{a}^{\dagger k} \hat{a}^l \rangle$
with $k$ or $l$ (or both) being equal to $2^d$.

\subsection{Detection of $\langle :\hat{x}_\varphi^k \hat{p}_\varphi^l: \rangle$}

\begin{figure}
    \includegraphics{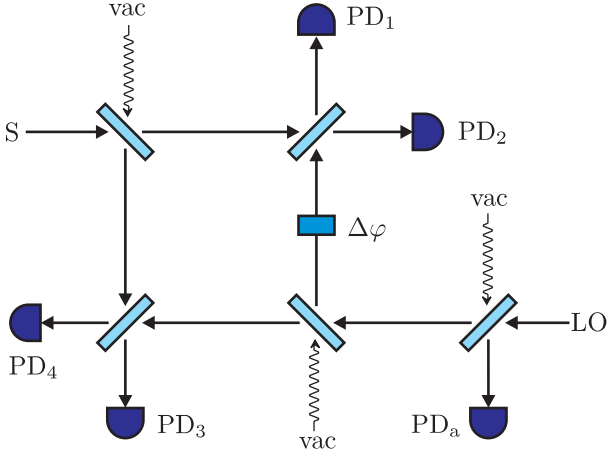}
\caption{Scheme for measuring the moments $\langle:\hat{x}_\varphi^k\hat{p}_\varphi^l:\rangle$.}\label{fig:xp}
\end{figure}

The moments $\langle :\hat{x}_\varphi^k \hat{p}_\varphi^l: \rangle$ can be measured
with the scheme presented on the Fig.~\ref{fig:xp}, as it has been
proposed in~\cite{pra-71-011802}. In fact, this is a slightly modified version of the Noh-Foug\`{e}res-Mandel device~\cite{prl-67-1426}. The lowest-order moments are
explicitely expressed in terms of the correlation functions
$\Gamma_j$, $\Gamma_{jk}$ according to the following formulas:
\begin{equation}
    \langle \hat{x}_\varphi \rangle = \sqrt{2} \frac{\Gamma_3 - \Gamma_4}{|\alpha|}, \quad
    \langle \hat{p}_\varphi \rangle = \sqrt{2} \frac{\Gamma_1 - \Gamma_2}{|\alpha|}.
\end{equation}
The second-order moments are of the form
\begin{equation}
\begin{split}
    \langle : \hat{x}_\varphi^2 : \rangle &= 4 \frac{\Gamma_{12} - \Gamma_{34}}{|\alpha|^2} +
    \langle \hat{n} \rangle, \\
    \langle : \hat{p}_\varphi^2 : \rangle &= 4 \frac{\Gamma_{34} - \Gamma_{12}}{|\alpha|^2} +
    \langle \hat{n} \rangle, \\
    \langle : \hat{x}_\varphi\hat{p}_\varphi : \rangle &= 4 \frac{\Gamma_{13} + \Gamma_{24} - \Gamma_{12} -
    \Gamma_{34}}{|\alpha|^2} - \langle \hat{n} \rangle,
\end{split}
\end{equation}
where
\begin{equation}
    \langle \hat{n} \rangle = \Gamma_1 + \Gamma_2 + \Gamma_3 + \Gamma_4 - |\alpha|^2.
\end{equation}
The amplitude $|\alpha|$ of the local oscillator is detected by the auxiliary photodetector $\mathrm{PD}_\mathrm{a}$. Note that
one could also try to use the measured data more efficiently as
discussed in detail in the preceding subsection. However, this would
lead to somewhat more complex expressions for the moments. Since the
procedure is straightforward, we do not present it here.

The scheme in the form shown in Fig.~\ref{fig:xp} allows one to
determine all the moments $\langle :
\hat{x}^k_\varphi\hat{p}^l_\varphi : \rangle$ up to the order of
$k+l=4$. The further extension of the method is straightforward. For
example, each of the detectors $\mathrm{PD}_i$ ($i=1\ldots 4$) can be
replaced with a beam splitter, each of which mixing the field with a
vacuum input. In their output ports the output fields are detected
by pairs of photodetectors. This allows us to extract the moments up
to the order of $k+l=8$, and so forth.

\subsection{Detection of $\langle:\hat{x}_\varphi^k\hat{n}^l:\rangle$}

\begin{figure}
    \includegraphics{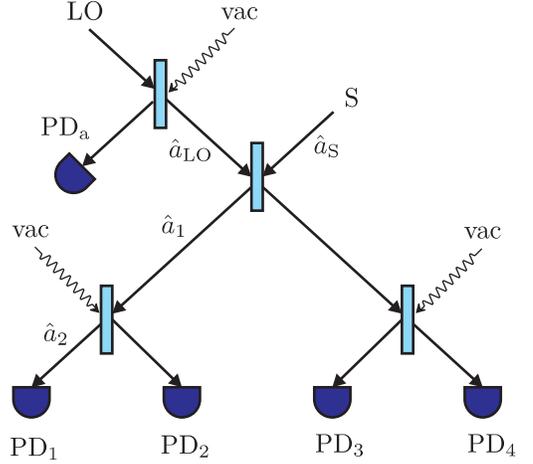}
\caption{Scheme for measuring the moments $\langle:\hat{x}_\varphi^k\hat{n}^l:\rangle$.}\label{fig:moments2}
\end{figure}

Figure~\ref{fig:moments2} illustrates the possibility of measuring
moments $\langle:\hat{x}_\varphi^k\hat{n}^l:\rangle$ by an extended version of the homodyne cross correlation scheme considered in Ref.~\cite{pra-51-4160}.  The moments
$\langle\hat{n}\rangle$ and $\langle\hat{x}_\varphi\rangle$ can be
obtained according to the following relations
\begin{equation}
\begin{split}
    \langle\hat{n}\rangle &= \Gamma_1 + \Gamma_2 + \Gamma_3 + \Gamma_4 - |\alpha|^2, \\
    \langle\hat{x}_\varphi\rangle &= \frac{1}{|\alpha|^2}\Bigl(\Gamma_1 + \Gamma_2 - \Gamma_3 -\Gamma_4\Bigr).
\end{split}
\end{equation}
Higher-order moments can be extracted from the measured data step by
 step.
 
 We give explicit expressions only for the second-order moments.  The
 moment $\langle:\hat{n}^2:\rangle$ can be directly measured with
 blocked local oscillator.  Note that this moment and the first-order
 moments together allow one to calculate the moment
 $\langle:\hat{x}_\varphi^2:\rangle$:
\begin{equation}
\begin{split}
    \Gamma_{13} &+ \Gamma_{14} + \Gamma_{23} + \Gamma_{24} = \\
&\frac{1}{4} \Bigl(\langle:\hat{n}^2:\rangle + 2|\alpha|^2\langle\hat{n}\rangle + |\alpha|^4 - |\alpha|^2\langle:\hat{x}_\varphi^2:\rangle\Bigr).
\end{split}
\end{equation}
The moment $\langle:\hat{n}\hat{x}_\varphi:\rangle$ can be obtained from the simple relation
\begin{equation}
    \Gamma_{12} - \Gamma_{34} =
\frac{1}{4}\Bigl(|\alpha|\langle:\hat{n}\hat{x}_\varphi:\rangle +
|\alpha|^2\langle\hat{x}_\varphi\rangle\Bigr).
\end{equation}
All higher-order moments can be measured in such a way with an
appropriately extended version of the scheme shown in Fig.~\ref{fig:moments2}.

\section{Amplitude-squared squeezing}

In this section we will consider a particular nonclassical effect
which can be well described in terms of the moments-based
nonclassicality criteria derived in Sec.~\ref{sec-II}. The examples
will also be simple from the viewpoint of the measurement of the
needed moments, when the measurement principles of Sec.~\ref{sec-III}
are used. This example is the amplitude squared squeezing~\cite{pra-36-3796}. We note that for the measurement of
amplitude-squared squeezing to our knowledge there exists no direct
measurement principle.  It has been proposed to measure the effect
after the interaction with a Kerr medium~\cite{pra-44-4578}. Such
techniques are, however, not easy to use and require sufficiently
strong signal fields for realizing the nonlinear interaction before
the detection. Hence a possibility of a more direct measurement of the
effect would be of interest.

Let us start with the characterization of the nonclassical effects we
are interested in. As a special choice of $\hat{f}$ in the
nonclassicality condition~\eqref{eq:ff2} let us consider
the following operator
\begin{equation}
    \hat{X}_\varphi = \hat{a}^2 e^{i\varphi} + \hat{a}^{\dagger 2} e^{-i\varphi}.
\end{equation}
According to \eqref{eq:ff} for any number $c$ the mean value
\begin{equation}
    \langle : (\hat{X}_\varphi-c)^2 : \rangle \geqslant 0
\end{equation}
is nonnegative for all classical states. For $c = \langle\hat{X}_\varphi\rangle$ this gives the following condition:
\begin{equation}
    \langle : (\Delta\hat{X}_\varphi)^2 : \rangle \geqslant 0,
\end{equation}
or, in other words, the condition
\begin{equation}
    \langle : (\Delta\hat{X}_\varphi)^2 : \rangle < 0
\end{equation}
is sufficient for nonclassicality. We note that the condition
for amplitude-squared squeezing can be easily expressed as
\begin{equation}
    s^{(2)}_\varphi =
    \begin{vmatrix}
      1 & \langle:\hat{x}_\varphi\hat{p}_\varphi:\rangle \\[0.5ex]
      \langle:\hat{x}_\varphi\hat{p}_\varphi:\rangle &
      \langle:\hat{x}_\varphi^2\hat{p}^2_\varphi:\rangle &
\end{vmatrix} < 0,
\end{equation}
in terms of the moments of two quadratures.

Another way to formulate the condition for amplitude-squared squeezing is based on the moments of the operators
$\hat{a}^\dagger$, $\hat{a}$. Let us consider the subdeterminant
$s_3$ that results from the determinant \eqref{eq:daa} by
canceling all its rows and columns except those beginning with the
elements $1$,
$\langle\hat{a}^2\rangle$ and
$\langle\hat{a}^{\dagger 2}\rangle$. This leads to a nonclassicality condition of the form
\begin{equation}\label{eq:dk}
    s_3 =
\begin{vmatrix}
    1 & \langle\hat{a}^{\dagger 2}\rangle & \langle\hat{a}^2\rangle \\
    \langle\hat{a}^2\rangle & \langle\hat{a}^{\dagger 2}\hat{a}^2\rangle &
    \langle\hat{a}^4\rangle \\
    \langle\hat{a}^{\dagger 2}\rangle & \langle\hat{a}^{\dagger 4}\rangle & \langle\hat{a}^{\dagger 2}\hat{a}^2\rangle
\end{vmatrix} < 0.
\end{equation}
We show that the negativity of this determinant is equivalent to the following condition
\begin{equation}\label{eq:X}
    \Bigl\langle:(\Delta \hat{X}_\varphi)^2:\Bigr\rangle
    < 0, \quad \forall\varphi.
\end{equation}
In fact, the determinant \eqref{eq:dk} can be written as follows:
\begin{equation}\label{eq:dmin-max}
    s_3 = \frac{1}{4} \min_\varphi \Bigl\langle:(\Delta \hat{X}_\varphi)^2:\Bigr\rangle \max_\varphi \Bigl\langle:(\Delta \hat{X}_\varphi)^2:\Bigr\rangle.
\end{equation}
It is not difficult to see that the last term in the product on the right hand side of this equality is always nonnegative
\begin{equation}
    \max_\varphi \Bigl\langle:(\Delta \hat{X}_\varphi)^2:\Bigr\rangle \geqslant 0.
\end{equation}
The explicit expressions for the minimum and the maximum of the variance of the operator $\hat{X}_\varphi$ are
\begin{equation}\label{eq:min-max}
\begin{split}
  \min_\varphi \Bigl\langle:(\Delta \hat{X}_\varphi)^2:\Bigr\rangle &=
  2\Bigl[\Bigl\langle\Delta\hat{a}^{\dagger 2}\Delta\hat{a}^2\Bigr\rangle-
  \Bigl|\Bigl\langle(\Delta\hat{a}^2)^2\Bigr\rangle\Bigr|\Bigr], \\
  \max_\varphi \Bigl\langle:(\Delta
  \hat{X}_\varphi)^2:\Bigr\rangle &=
  2\Bigl[\Bigl\langle\Delta\hat{a}^{\dagger 2}\Delta\hat{a}^2\Bigr\rangle+
  \Bigl|\Bigl\langle(\Delta\hat{a}^2)^2\Bigr\rangle\Bigr|\Bigr].
\end{split}
\end{equation}
It is clear that the mean value $\Bigl\langle\Delta\hat{a}^{\dagger 2}\Delta\hat{a}^2\Bigr\rangle$
is always nonnegative. Due to the second of the equalities~\eqref{eq:min-max} the maximal variance
of $\hat{X}_\varphi$ is always nonnegative. Hence the condition for the negativity of $s_3$ in
Eq.~\eqref{eq:dmin-max} is a direct demonstration of amplitude-squared squeezing.

An advantage of this method for demonstrating amplitude squared
squeezing consists in the fact that we do not need to adjust the local
oscillator phase to the noise minimum. The negativity of the
considered subdeterminant demonstrates the negativity of the minimum
of the normally-ordered variance already. Moreover, the moments $\langle\hat{a}^2\rangle$, $\langle\hat{a}^4\rangle$ and $\langle\hat{a}^{\dagger 2} \hat{a}^2\rangle$ needed in the condition \eqref{eq:dk} are easily obtained by the methods of the preceding section.

\section{minimum-uncertainty amplitude-squared squeezed states}

Let us consider amplitude squared-squeezed states in more
detail (\cite{pra-36-3796, pra-43-515, qo-6-37}). By definition, a state is
amplitude-squared squeezed if
\begin{equation}
    \min_\varphi \langle : (\Delta\hat{X}_\varphi)^2 :
\rangle < 0,
\end{equation}
which can be rewritten in the equivalent form as
\begin{equation}\label{eq:Xn}
    \exists \varphi: \quad \langle (\Delta\hat{X}_\varphi)^2 \rangle <
4 \langle\hat{n}\rangle +2.
\end{equation}
We consider here a special class of amplitude-squared squeezed states that satisfy the minimum
uncertainty relation~\cite{pra-43-515}. For this purpose we introduce the operators
\begin{equation}
    \hat{X} = \hat{X}_\varphi, \quad
    \hat{Y} = \hat{X}_{\varphi+\pi/2}.
\end{equation}
They fulfill the uncertainty relation
\begin{equation}
    \langle(\Delta \hat{X})^2\rangle^{1/2}
    \langle(\Delta \hat{Y})^2\rangle^{1/2} \geqslant
    4 \langle\hat{n}\rangle +2.
\end{equation}
A state satisfies the minimum uncertainty relation if
\begin{equation}\label{eq:mur}
    \langle(\Delta \hat{X})^2\rangle^{1/2}
    \langle(\Delta \hat{Y})^2\rangle^{1/2} =
    4 \langle\hat{n}\rangle +2.
\end{equation}
In the following we will only consider pure quantum states.

As it has been shown in~\cite{qo-6-37}, a pure state $|\psi\rangle$
satisfies the condition \eqref{eq:mur} if it is a solution of the eigenvalue problem
\begin{equation}\label{eq:ep}
    (\hat{X} + i \lambda \hat{Y})|\psi\rangle = \beta |\psi\rangle,
\end{equation}
for all real nonnegative $\lambda$ and any complex $\beta$. One can easily check that form
Eq.~\eqref{eq:ep} it follows that
\begin{equation}\label{eq:XY}
\begin{split}
    \langle(\Delta \hat{X})^2\rangle &= \lambda
    (4 \langle\hat{n}\rangle +2), \\
    \langle(\Delta \hat{Y})^2\rangle &= \frac{1}{\lambda}
    (4 \langle\hat{n}\rangle +2).
\end{split}
\end{equation}
For either $\lambda>1$ or $1/\lambda > 1$ one of the variations on the
left-hand sides of Eqs~\eqref{eq:XY} satisfies the condition \eqref{eq:Xn}.
Therefore the state~\eqref{eq:ep} is always amplitude-square squeezed, except
for  $\lambda=1$.

\begin{figure}
    \subfigure[$\lambda=1.01$]{\includegraphics[width=38mm]{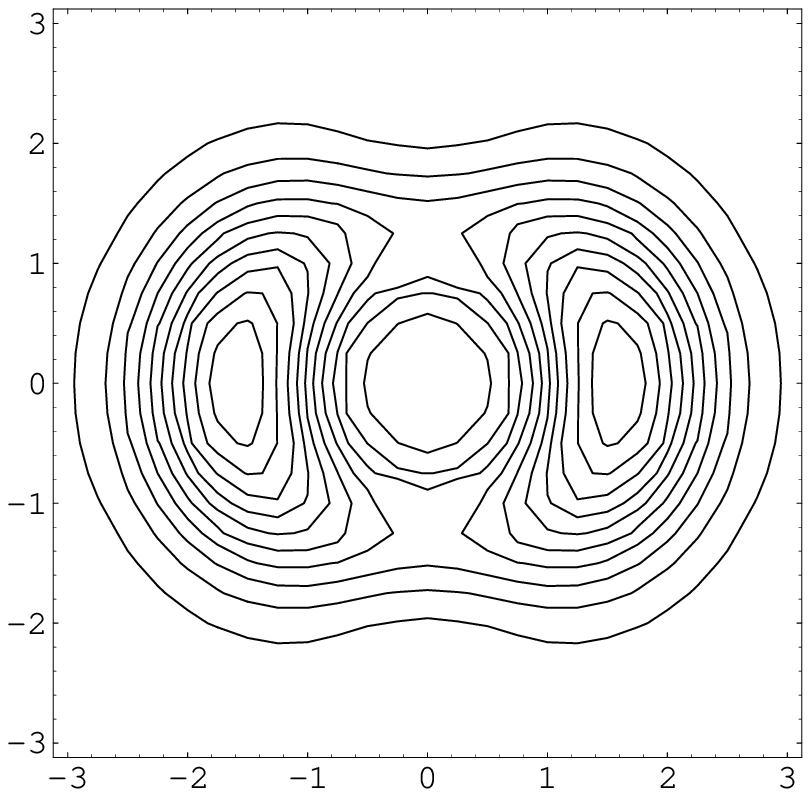}}
    \subfigure[$\lambda=1.05$]{\includegraphics[width=38mm]{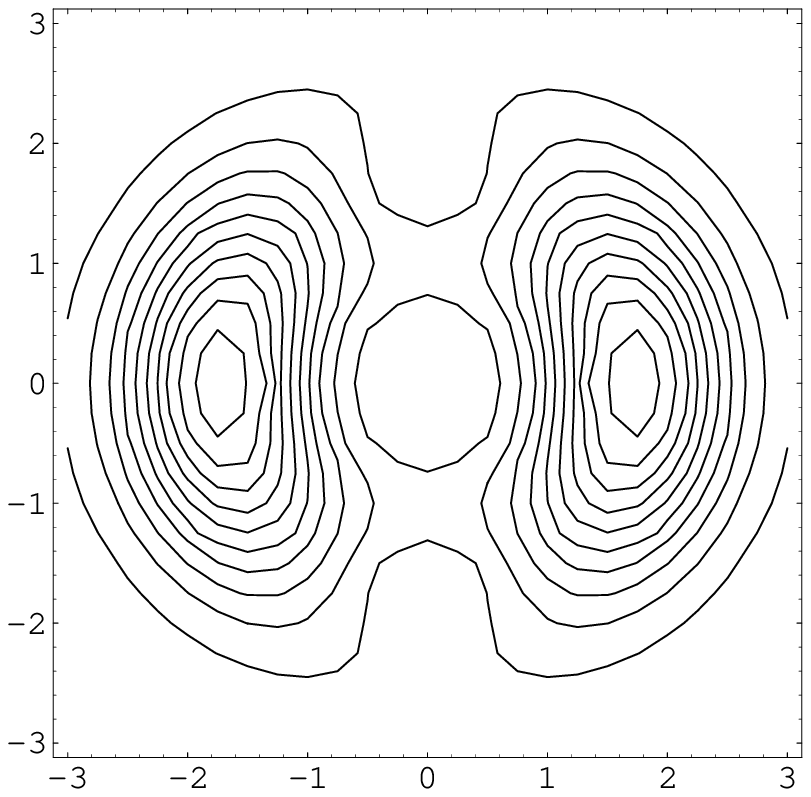}}
    \subfigure[$\lambda=1.1$]{\includegraphics[width=38mm]{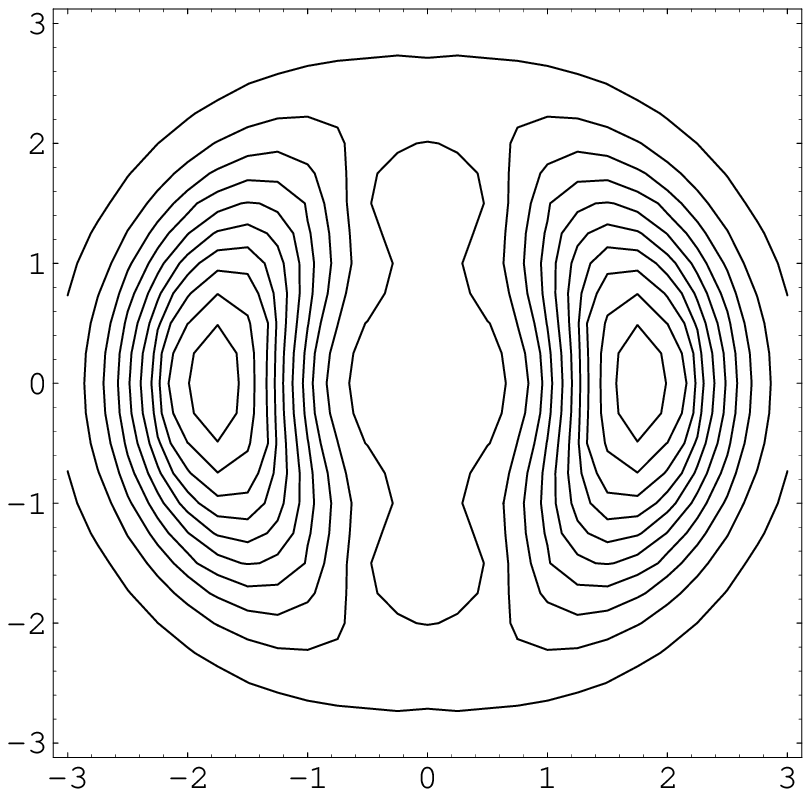}}
    \subfigure[$\lambda=2$]{\includegraphics[width=38mm]{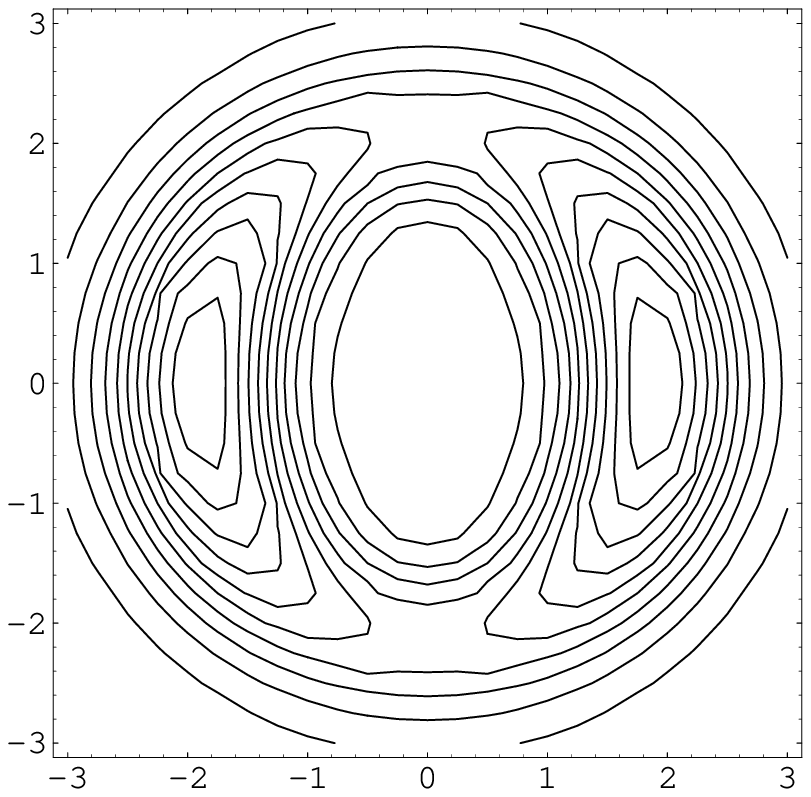}}
\caption{The $Q$ function of amplitude square squeezed states \eqref{eq:ass} for $m = 2$.}\label{fig:Qass}
\end{figure}

It was shown in \cite{pra-43-515} that there exist solutions of
Eq.~\eqref{eq:ep} of the form
\begin{equation}\label{eq:ass}
    |\psi(m, \lambda)\rangle = c_m(\lambda)\hat{S}(z)H_m(i \gamma\hat{a}^\dagger) |0\rangle,
\end{equation}
where
\begin{equation}\label{eq:gamma}
    \gamma = \gamma(\lambda) =
    \begin{cases}
        e^{i \pi/4}\sqrt{\sqrt{1-\lambda^2}/2} & 0 < \lambda < 1, \\
        \sqrt{\sqrt{\lambda^2 - 1}/{2\lambda}} & 1 < \lambda,
    \end{cases}
\end{equation}
and
\begin{equation}
    |c_m(\lambda)|^2 =
\begin{cases}
    1 & m = 0, \\
    \frac{(-1)^m}{m! C^{-m}_m(2|\gamma(\lambda)|^2)} & m > 0.
\end{cases}
\end{equation}
The parameter $\beta$ in the equation \eqref{eq:ep} is connected with $m$ and $\lambda$ by the expression
\begin{equation}
    \beta =
\begin{cases}
    i \sqrt{1-\lambda^2}(2m + 1) & 0 < \lambda < 1, \\
    \sqrt{\lambda^2-1}(2m+1) & 1 < \lambda.
\end{cases}
\end{equation}
The parameter $z = z(\lambda) = r e^{i \varphi}$ reads
\begin{equation}\label{eq:z}
    \tanh^2r = \frac{\lambda-1}{\lambda+1}e^{-2i\varphi}, \quad
    \varphi =
\begin{cases}
    \pi/2 & 0 < \lambda < 1, \\
    0 & 1 < \lambda.
\end{cases}
\end{equation}
Some examples of the $Q$-function of the states~\eqref{eq:ass} are shown in Fig.~\ref{fig:Qass}.

\begin{figure}
    \includegraphics{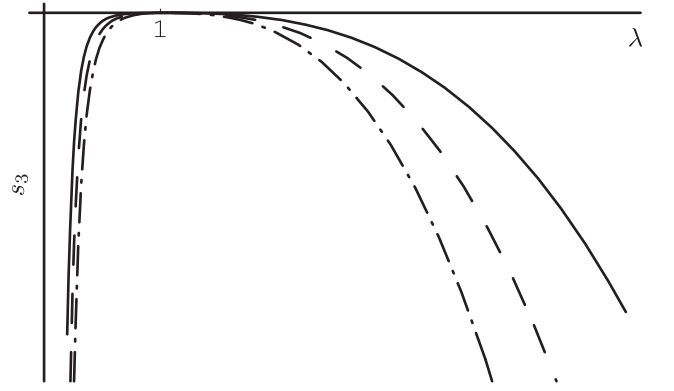}
\caption{The determinants $s_3$ \eqref{eq:dk} for $m=2$ (solid), $m=3$ (dashed) and $m=4$
(dash-dotted).}\label{fig:d2}
\end{figure}

In Fig.~\ref{fig:d2} we illustrate examples of the determinants for the
minimum-uncertainty amplitude-squared squeezed states under study.  It is of
great importance that all the moments appearing in the nonclassicality
condition~\eqref{eq:dk} can be determined by the homodyne correlation
measurement technique proposed in Sec.~\ref{subsec-IIIa}. Unless the methods
of quantum-state reconstruction, for a review cf.~\cite{welsch}, the
measurements proposed here are even possible in cases when the
overall quantum efficiency of the detection device is small. 

\section{Summary and Conclusion}

We have shown that condition for the nonclassicality of a single-mode quantum
state, that is the failure of the $P$-function to be a probability measure,
can be formulated in different ways. All the versions of necessary and
sufficient conditions the we have formulated in this paper turn out to be
special representations of the violation of Bochner's condition for the
existence of a classical characteristic function of the $P$-function. The
considered formulations of different forms of nonclassicality criteria include
characteristic functions of quadratures and different types of
normally-ordered moments.  Most importantly, all the quantities under study
are accessible to observation.  Their measurement, however, requires to
develop new types of measurement principles.

We have proposed rather simple and direct methods of measuring three kinds of
normally-ordered moments of a single-mode radiation field. For each type of
moments a particular detection scheme is analyzed: for the moments $\langle
\hat{a}^{\dagger k} \hat{a}^l \rangle$ of the creation and annihilation
operators, the moments $\langle
:\hat{x}^k_\varphi \hat{p}^l_\varphi: \rangle$ of two quadratures
and the moments
$\langle:\hat{x}^k_\varphi\hat{n}^l:\rangle$ of a quadrature 
and the photon number operator.
In all the considered schemes the total number $N$ of photo-detectors needed to measure these moments
is at most twice as large as the largest
value of $k$ and $l$: $N < 2 \max(k, l)$. 
We also tried to present the extraction procedure for the moments in an 
optimal way form the viewpoint of an effective use of the available measured
data. This means that one tries to use all the data obtained in measurements. 
Unfortunately this does not provide the expressions for the moments of
interest in its simplest form. In the simplest form one would 
not need to sum up the correlation functions of all possible
combinations of photo-detectors to get the moments. However, one would lose
part of the measured data and thus this simplified approach would be less
precise from the experimental point of view. 

Our new methods of characterizing nonclassical effects by moments of
annihilation and creation operators are applied to the characterization of
amplitude-squared squeezing. We have shown that our criteria give a direct
insight in the effect of amplitude-squared squeezing without the need of
adjusting the phase of the local oscillator. More importantly, the
demonstration of amplitude-squared squeezing until now was considered to
require a rather difficult and indirect observation procedure, so that this
effect received little attention in the context of experiments. The
measurement procedures proposed in this paper turn out to be of
particular simplicity for demonstrating the amplitude-squared squeezing
effect. This may lead to new interest in this special nonclassical effect.

In conclusion we have formulated new types of criteria for characterizing
nonclassical effects of radiation fields. For all versions of nonclassicality
criteria we have proposed appropriate and simple measurement principles. These principles are based on homodyne correlation techniques with a
weak local oscillator. They can be used even when the quantum efficiency of
the device is small. The proposed methods may open new possibilities of
demonstration and practical application of nonclassical effects of radiation fields.

\begin{acknowledgments}
The authors gratefully acknowledge valuable discussions with R. Kn\"orr, F. Liese and Th. Richter.
\end{acknowledgments}

\appendix*

\section{Moments of amplitude-squared squeezed states}\label{app:A}

The moments $\langle \hat{a}^{\dagger k} \hat{a}^l \rangle_m$ of the state $|\psi_m\rangle$ can be
obtained in the following way. Let us take
\begin{equation}
    |\tilde{\psi}_m\rangle = c^{-1}_m |\psi_m\rangle = \hat{S}(z) H_m(i \gamma
    \hat{a}^\dagger)|0\rangle,
\end{equation}
where the parameters $z$ and $\gamma$ are defined by the Eqs.~\eqref{eq:z} and \eqref{eq:gamma}
correspondingly. It is clear that
\begin{equation}
    \langle \hat{a}^{\dagger k} \hat{a}^l \rangle_m = \langle \tilde{\psi}_m|
    \hat{a}^{\dagger k} \hat{a}^l |\tilde{\psi}_m\rangle.
\end{equation}
We calculate the quantities $\langle \tilde{\psi}_m| \hat{a}^{\dagger k} \hat{a}^l
|\tilde{\psi}_m\rangle$ with the help of the generating function
\begin{equation}
    F(x, y, u, v) = \sum^{+\infty}_{n, m, k, l = 0} \langle \tilde{\psi}_n|
    \hat{a}^{\dagger k} \hat{a}^l |\tilde{\psi}_m\rangle \frac{x^n y^m u^k v^l}{n! m! k! l!}.
\end{equation}
Clearly,
\begin{equation}
\begin{split}
    \sum^{+\infty}_{n, m = 0} &\langle \tilde{\psi}_n|
    \hat{a}^{\dagger k} \hat{a}^l |\tilde{\psi}_m\rangle \frac{x^n y^m}{n! m!} = \\
    &\langle 0|e^{-x^2-2i\overline{\gamma}x\hat{a}}\hat{S}^\dagger(z)\hat{a}^{\dagger k} \hat{a}^l
    \hat{S}(z)e^{-y^2+2i\gamma y \hat{a}^\dagger}|0\rangle,
\end{split}
\end{equation}
and the function $F(x, y, u, v)$ reads as
\begin{equation}
\begin{split}
    &F(x, y, u, v) = \\
    &\langle 0|e^{-x^2-2i\overline{\gamma}x\hat{a}}\hat{S}^\dagger(z)e^{u\hat{a}^\dagger}
    e^{v \hat{a}} \hat{S}(z)e^{-y^2+2i\gamma y \hat{a}^\dagger}|0\rangle.
\end{split}
\end{equation}
Using the relation
\begin{equation}
    \hat{a}\hat{S}(z) = \hat{S}(z)(\mu \hat{a} + \nu \hat{a}^\dagger)
\end{equation}
it can be further simplified to the following form
\begin{equation}
    F(x, y, u, v) = \exp\Bigl(-\frac{1}{2}\vec{w}^T A \vec{w}\Bigr),
\end{equation}
where $\vec{w} = (x, y, u, v)^T$ and
\begin{equation}
    A =
    \begin{pmatrix}
        2 & -4|\gamma|^2 & -2i\overline{\nu}\gamma & -2i\mu\gamma \\
        -4|\gamma|^2 & 2 & 2i\mu\overline{\gamma} & 2i\nu\overline{\gamma} \\
        -2i\overline{\nu}\gamma & 2i\mu\overline{\gamma} & -\mu\overline{\nu} & -|\nu|^2 \\
        -2i\mu\gamma & 2i\nu\overline{\gamma} & -|\nu|^2 & -\mu\nu
    \end{pmatrix}.
\end{equation}
From this expression it follows that
\begin{equation}
    \langle \tilde{\psi}_n| \hat{a}^{\dagger k} \hat{a}^l |\tilde{\psi}_m\rangle
    = H^{\{A\}}_{nmkl}(0, 0, 0, 0),
\end{equation}
where $H^{\{A\}}_{nmkl}(0, 0, 0, 0)$ is a four-dimensional  Hermite polynomial defined as ($\vec{e}
= (a, b, c, d)^T$)
\begin{equation}
\begin{split}
    \sum^{+\infty}_{n, m, k, l = 0} &H^{\{A\}}_{nmkl}(a, b, c, d) \frac{x^n y^m u^k v^l}{n! m! k!
    l!} = \\
    &\exp\left(-\frac{1}{2}\vec{w}^T A \vec{w} + \vec{w}^T A \vec{e}\right).
\end{split}
\end{equation}
Finally, we arrive at the result
\begin{equation}
    \langle \hat{a}^{\dagger k} \hat{a}^l \rangle_m = |c_m|^2 H^{\{A\}}_{mmkl}(0, 0, 0, 0),
\end{equation}
for the moments we are interested in.


\begin{thebibliography}{32}
\expandafter\ifx\csname natexlab\endcsname\relax\def\natexlab#1{#1}\fi
\expandafter\ifx\csname bibnamefont\endcsname\relax
  \def\bibnamefont#1{#1}\fi
\expandafter\ifx\csname bibfnamefont\endcsname\relax
  \def\bibfnamefont#1{#1}\fi
\expandafter\ifx\csname citenamefont\endcsname\relax
  \def\citenamefont#1{#1}\fi
\expandafter\ifx\csname url\endcsname\relax
  \def\url#1{\texttt{#1}}\fi
\expandafter\ifx\csname urlprefix\endcsname\relax\def\urlprefix{URL }\fi
\providecommand{\bibinfo}[2]{#2}
\providecommand{\eprint}[2][]{\url{#2}}

\bibitem[{\citenamefont{Kimble et~al.}(1977)\citenamefont{Kimble, Dagenais, and
  Mandel}}]{prl-39-691}
\bibinfo{author}{\bibfnamefont{H.~J.} \bibnamefont{Kimble}},
  \bibinfo{author}{\bibfnamefont{M.}~\bibnamefont{Dagenais}}, \bibnamefont{and}
  \bibinfo{author}{\bibfnamefont{L.}~\bibnamefont{Mandel}},
  \bibinfo{journal}{Phys. Rev. Lett.} \textbf{\bibinfo{volume}{39}},
  \bibinfo{pages}{691} (\bibinfo{year}{1977}).

\bibitem[{\citenamefont{Short and Mandel}(1983)}]{prl-51-384}
\bibinfo{author}{\bibfnamefont{R.}~\bibnamefont{Short}} \bibnamefont{and}
  \bibinfo{author}{\bibfnamefont{L.}~\bibnamefont{Mandel}},
  \bibinfo{journal}{Phys. Rev. Lett.} \textbf{\bibinfo{volume}{51}},
  \bibinfo{pages}{384} (\bibinfo{year}{1983}).

\bibitem[{\citenamefont{Slusher et~al.}(1985)\citenamefont{Slusher, Hollberg,
  Yurke, Mertz, and Valley}}]{prl-55-2409}
\bibinfo{author}{\bibfnamefont{R.}~\bibnamefont{Slusher}},
  \bibinfo{author}{\bibfnamefont{L.}~\bibnamefont{Hollberg}},
  \bibinfo{author}{\bibfnamefont{B.}~\bibnamefont{Yurke}},
  \bibinfo{author}{\bibfnamefont{J.}~\bibnamefont{Mertz}}, \bibnamefont{and}
  \bibinfo{author}{\bibfnamefont{J.}~\bibnamefont{Valley}},
  \bibinfo{journal}{Phys. Rev. Lett.} \textbf{\bibinfo{volume}{55}},
  \bibinfo{pages}{2409} (\bibinfo{year}{1985}).

\bibitem[{\citenamefont{Smithey et~al.}(1993)\citenamefont{Smithey, Beck,
  Raymer, and Faridani}}]{prl-70-1244}
\bibinfo{author}{\bibfnamefont{D.~T.} \bibnamefont{Smithey}},
  \bibinfo{author}{\bibfnamefont{M.}~\bibnamefont{Beck}},
  \bibinfo{author}{\bibfnamefont{M.~G.} \bibnamefont{Raymer}},
  \bibnamefont{and} \bibinfo{author}{\bibfnamefont{A.}~\bibnamefont{Faridani}},
  \bibinfo{journal}{Phys. Rev. Lett.} \textbf{\bibinfo{volume}{70}},
  \bibinfo{pages}{1244} (\bibinfo{year}{1993}).

\bibitem[{\citenamefont{Welsch et~al.}(1999)\citenamefont{Welsch, Vogel, and
  Opatrn\'{y}}}]{welsch}
\bibinfo{author}{\bibfnamefont{D.-G.} \bibnamefont{Welsch}},
  \bibinfo{author}{\bibfnamefont{W.}~\bibnamefont{Vogel}}, \bibnamefont{and}
  \bibinfo{author}{\bibfnamefont{T.}~\bibnamefont{Opatrn\'{y}}},
  \emph{\bibinfo{title}{Homodyne detection and quantum-state reconstraction}}
  (\bibinfo{publisher}{Elsevier science B. V.}, \bibinfo{year}{1999}),
  vol.~\bibinfo{volume}{39} of \emph{\bibinfo{series}{Progress in Optics}},
  chap.~\bibinfo{chapter}{2}, pp. \bibinfo{pages}{63--211}.

\bibitem[{\citenamefont{Vogel}(1991)}]{prl-67-2450}
\bibinfo{author}{\bibfnamefont{W.}~\bibnamefont{Vogel}},
  \bibinfo{journal}{Phys. Rev. Lett.} \textbf{\bibinfo{volume}{67}},
  \bibinfo{pages}{2450} (\bibinfo{year}{1991}).

\bibitem[{\citenamefont{Vogel}(1995)}]{pra-51-4160}
\bibinfo{author}{\bibfnamefont{W.}~\bibnamefont{Vogel}},
  \bibinfo{journal}{Phys. Rev. A} \textbf{\bibinfo{volume}{51}},
  \bibinfo{pages}{4160} (\bibinfo{year}{1995}).

\bibitem[{\citenamefont{Carmichael et~al.}(2000)\citenamefont{Carmichael,
  Castro-Beltran, Foster, and Orozco}}]{prl-85-1855}
\bibinfo{author}{\bibfnamefont{H.~J.} \bibnamefont{Carmichael}},
  \bibinfo{author}{\bibfnamefont{H.~M.} \bibnamefont{Castro-Beltran}},
  \bibinfo{author}{\bibfnamefont{G.~T.} \bibnamefont{Foster}},
  \bibnamefont{and} \bibinfo{author}{\bibfnamefont{L.~A.}
  \bibnamefont{Orozco}}, \bibinfo{journal}{Phys. Rev. Lett.}
  \textbf{\bibinfo{volume}{85}}, \bibinfo{pages}{1855} (\bibinfo{year}{2000}).

\bibitem[{\citenamefont{Beck et~al.}(2001)\citenamefont{Beck, Dorrer, and
  Walmsley}}]{prl-87-253601}
\bibinfo{author}{\bibfnamefont{M.}~\bibnamefont{Beck}},
  \bibinfo{author}{\bibfnamefont{C.}~\bibnamefont{Dorrer}}, \bibnamefont{and}
  \bibinfo{author}{\bibfnamefont{I.~A.} \bibnamefont{Walmsley}},
  \bibinfo{journal}{Phys. Rev. Lett.} \textbf{\bibinfo{volume}{87}},
  \bibinfo{pages}{253601} (\bibinfo{year}{2001}).

\bibitem[{\citenamefont{Titulaer and Glauber}(1965)}]{pr-140B-676}
\bibinfo{author}{\bibfnamefont{U.}~\bibnamefont{Titulaer}} \bibnamefont{and}
  \bibinfo{author}{\bibfnamefont{R.}~\bibnamefont{Glauber}},
  \bibinfo{journal}{Phys. Rev.} \textbf{\bibinfo{volume}{140}},
  \bibinfo{pages}{B676} (\bibinfo{year}{1965}).

\bibitem[{\citenamefont{Mandel}(1986)}]{scr-T12-34}
\bibinfo{author}{\bibfnamefont{L.}~\bibnamefont{Mandel}},
  \bibinfo{journal}{Phys. Scr.} \textbf{\bibinfo{volume}{T12}},
  \bibinfo{pages}{34} (\bibinfo{year}{1986}).

\bibitem[{\citenamefont{Vogel}(2000)}]{prl-84-1849}
\bibinfo{author}{\bibfnamefont{W.}~\bibnamefont{Vogel}},
  \bibinfo{journal}{Phys. Rev. Lett.} \textbf{\bibinfo{volume}{84}},
  \bibinfo{pages}{1849} (\bibinfo{year}{2000}).

\bibitem[{\citenamefont{Johansen}(2004)}]{pla-329-184}
\bibinfo{author}{\bibfnamefont{L.~M.} \bibnamefont{Johansen}},
  \bibinfo{journal}{Phys. Lett. A} \textbf{\bibinfo{volume}{329}},
  \bibinfo{pages}{184} (\bibinfo{year}{2004}).

\bibitem[{\citenamefont{Johansen and Luis}(2004)}]{pra-70-052115}
\bibinfo{author}{\bibfnamefont{L.~M.} \bibnamefont{Johansen}} \bibnamefont{and}
  \bibinfo{author}{\bibfnamefont{A.}~\bibnamefont{Luis}},
  \bibinfo{journal}{Phys. Rev. A} \textbf{\bibinfo{volume}{70}},
  \bibinfo{pages}{052115} (\bibinfo{year}{2004}).

\bibitem[{\citenamefont{Lvovsky and Shapiro}(2002)}]{pra-65-033830}
\bibinfo{author}{\bibfnamefont{A.}~\bibnamefont{Lvovsky}} \bibnamefont{and}
  \bibinfo{author}{\bibfnamefont{J.~H.} \bibnamefont{Shapiro}},
  \bibinfo{journal}{Phys. Rev. A} \textbf{\bibinfo{volume}{65}},
  \bibinfo{pages}{033830} (\bibinfo{year}{2002}).

\bibitem[{\citenamefont{Richter and Vogel}(2002)}]{prl-89-283601}
\bibinfo{author}{\bibfnamefont{T.}~\bibnamefont{Richter}} \bibnamefont{and}
  \bibinfo{author}{\bibfnamefont{W.}~\bibnamefont{Vogel}},
  \bibinfo{journal}{Phys. Rev. Lett.} \textbf{\bibinfo{volume}{89}},
  \bibinfo{pages}{283601} (\bibinfo{year}{2002}).

\bibitem[{\citenamefont{Bochner}(1933)}]{ma-108-378}
\bibinfo{author}{\bibfnamefont{S.}~\bibnamefont{Bochner}},
  \bibinfo{journal}{Math. Ann.} \textbf{\bibinfo{volume}{108}},
  \bibinfo{pages}{378} (\bibinfo{year}{1933}).

\bibitem[{\citenamefont{Kawata}(1972)}]{k-fapt}
\bibinfo{author}{\bibfnamefont{T.}~\bibnamefont{Kawata}},
  \emph{\bibinfo{title}{Fourier Analysis in Probability Theory}}
  (\bibinfo{publisher}{Academic Press}, \bibinfo{address}{New York},
  \bibinfo{year}{1972}).

\bibitem[{\citenamefont{Shchukin et~al.}(2005)\citenamefont{Shchukin, Richter,
  and Vogel}}]{pra-71-011802}
\bibinfo{author}{\bibfnamefont{E.}~\bibnamefont{Shchukin}},
  \bibinfo{author}{\bibfnamefont{T.}~\bibnamefont{Richter}}, \bibnamefont{and}
  \bibinfo{author}{\bibfnamefont{W.}~\bibnamefont{Vogel}},
  \bibinfo{journal}{Phys. Rev. A} \textbf{\bibinfo{volume}{71}},
  \bibinfo{pages}{011802} (\bibinfo{year}{2005}).

\bibitem[{\citenamefont{Agarwal and Tara}(1992)}]{pra-46-485}
\bibinfo{author}{\bibfnamefont{G.}~\bibnamefont{Agarwal}} \bibnamefont{and}
  \bibinfo{author}{\bibfnamefont{K.}~\bibnamefont{Tara}},
  \bibinfo{journal}{Phys. Rev. A} \textbf{\bibinfo{volume}{46}},
  \bibinfo{pages}{485} (\bibinfo{year}{1992}).

\bibitem[{\citenamefont{Agarwal}(1993)}]{oc-95-109}
\bibinfo{author}{\bibfnamefont{G.}~\bibnamefont{Agarwal}},
  \bibinfo{journal}{Opt. Commun.} \textbf{\bibinfo{volume}{95}},
  \bibinfo{pages}{109} (\bibinfo{year}{1993}).

\bibitem[{\citenamefont{Klyshko}(1996)}]{pu-39-573}
\bibinfo{author}{\bibfnamefont{D.~N.} \bibnamefont{Klyshko}},
  \bibinfo{journal}{Physics-Uspekhi} \textbf{\bibinfo{volume}{39}},
  \bibinfo{pages}{573} (\bibinfo{year}{1996}).

\bibitem[{\citenamefont{Korbicz et~al.}(2005)\citenamefont{Korbicz, Cirac,
  Wehr, and Lewenstein}}]{prl-94-153601}
\bibinfo{author}{\bibfnamefont{J.~K.} \bibnamefont{Korbicz}},
  \bibinfo{author}{\bibfnamefont{J.~I.} \bibnamefont{Cirac}},
  \bibinfo{author}{\bibfnamefont{J.}~\bibnamefont{Wehr}}, \bibnamefont{and}
  \bibinfo{author}{\bibfnamefont{M.}~\bibnamefont{Lewenstein}},
  \bibinfo{journal}{Phys. Rev. Lett.} \textbf{\bibinfo{volume}{94}},
  \bibinfo{pages}{153601} (\bibinfo{year}{2005}).

\bibitem[{\citenamefont{Asb\'{o}th et~al.}(2005)\citenamefont{Asb\'{o}th,
  Calsamiglia, and Ritsch}}]{prl-94-173602}
\bibinfo{author}{\bibfnamefont{J.~K.} \bibnamefont{Asb\'{o}th}},
  \bibinfo{author}{\bibfnamefont{J.}~\bibnamefont{Calsamiglia}},
  \bibnamefont{and} \bibinfo{author}{\bibfnamefont{H.}~\bibnamefont{Ritsch}},
  \bibinfo{journal}{Phys. Rev. Lett.} \textbf{\bibinfo{volume}{94}},
  \bibinfo{pages}{173602} (\bibinfo{year}{2005}).

\bibitem[{\citenamefont{Hillery}(1987)}]{pra-36-3796}
\bibinfo{author}{\bibfnamefont{M.}~\bibnamefont{Hillery}},
  \bibinfo{journal}{Phys. Rev. A} \textbf{\bibinfo{volume}{36}},
  \bibinfo{pages}{3796} (\bibinfo{year}{1987}).

\bibitem[{\citenamefont{Sudarshan}(1963)}]{prl-10-277}
\bibinfo{author}{\bibfnamefont{E.}~\bibnamefont{Sudarshan}},
  \bibinfo{journal}{Phys. Rev. Lett.} \textbf{\bibinfo{volume}{10}},
  \bibinfo{pages}{277} (\bibinfo{year}{1963}).

\bibitem[{\citenamefont{Glauber}(1963)}]{pr-131-2766}
\bibinfo{author}{\bibfnamefont{R.~J.} \bibnamefont{Glauber}},
  \bibinfo{journal}{Phys. Rev.} \textbf{\bibinfo{volume}{131}},
  \bibinfo{pages}{2766} (\bibinfo{year}{1963}).

\bibitem[{\citenamefont{Prestel and Delzell}(2001)}]{positive-polynomials}
\bibinfo{author}{\bibfnamefont{A.}~\bibnamefont{Prestel}} \bibnamefont{and}
  \bibinfo{author}{\bibfnamefont{C.~N.} \bibnamefont{Delzell}},
  \emph{\bibinfo{title}{Positive Polynomials: From Hilbert's 17th Problem to
  Real Algebra}} (\bibinfo{publisher}{Springer}, \bibinfo{address}{Berlin
  Heidelberg}, \bibinfo{year}{2001}).

\bibitem[{\citenamefont{Noh et~al.}(1991)\citenamefont{Noh, Foug\`{e}res, and
  Mandel}}]{prl-67-1426}
\bibinfo{author}{\bibfnamefont{J.~W.} \bibnamefont{Noh}},
  \bibinfo{author}{\bibfnamefont{A.}~\bibnamefont{Foug\`{e}res}},
  \bibnamefont{and} \bibinfo{author}{\bibfnamefont{L.}~\bibnamefont{Mandel}},
  \bibinfo{journal}{Phys. Rev. Lett.} \textbf{\bibinfo{volume}{67}},
  \bibinfo{pages}{1426} (\bibinfo{year}{1991}).

\bibitem[{\citenamefont{Hillery}(1991)}]{pra-44-4578}
\bibinfo{author}{\bibfnamefont{M.}~\bibnamefont{Hillery}},
  \bibinfo{journal}{Phys. Rev. A} \textbf{\bibinfo{volume}{44}},
  \bibinfo{pages}{4578} (\bibinfo{year}{1991}).

\bibitem[{\citenamefont{Bergou et~al.}(1991)\citenamefont{Bergou, Hillery, and
  Yu}}]{pra-43-515}
\bibinfo{author}{\bibfnamefont{J.~A.} \bibnamefont{Bergou}},
  \bibinfo{author}{\bibfnamefont{M.}~\bibnamefont{Hillery}}, \bibnamefont{and}
  \bibinfo{author}{\bibfnamefont{D.}~\bibnamefont{Yu}}, \bibinfo{journal}{Phys.
  Rev. A} \textbf{\bibinfo{volume}{43}}, \bibinfo{pages}{515}
  (\bibinfo{year}{1991}).

\bibitem[{\citenamefont{Yu and Hillery}(1994)}]{qo-6-37}
\bibinfo{author}{\bibfnamefont{D.}~\bibnamefont{Yu}} \bibnamefont{and}
  \bibinfo{author}{\bibfnamefont{M.}~\bibnamefont{Hillery}},
  \bibinfo{journal}{Quantum. Opt.} \textbf{\bibinfo{volume}{6}},
  \bibinfo{pages}{37} (\bibinfo{year}{1994}).

\end{thebibliography}
\end{document}